\documentclass[acmtog,balance=false]{acmart}

\newcommand{\cell}{c}
\newcommand{\cellcorner}{\mathbf{u}}
\newcommand{\cellvert}{\mathbf{x}}
\newcommand{\cellverts}{\mathcal{V}}

\newcommand{\allquads}{\mathcal{Q}}
\newcommand{\sdf}{s}
\newcommand{\R}{\mathbb{R}}
\newcommand{\mesh}{\mathcal{M}}
\newcommand{\intCells}{\mathcal{C}}
\newcommand{\edge}{e}
\newcommand{\Hermitepoint}{\mathbf{h}}
\newcommand{\Hermitenormal}{\mathbf{n}}
\newcommand{\gradtrilinear}{\nabla_{\text{tri}}}
\newcommand{\centroid}{\mathbf{c}}
\newcommand{\fip}{\mathbf{p}}
\newcommand{\assigned}{\mathcal{A}}
\newcommand{\energy}{E}
\newcommand{\sdfenergy}{\energy_{d}}

\newcommand{\dcenergy}{\energy_{H}}
\newcommand{\Hermiteenergy}{\dcenergy}
\newcommand{\distance}{d}
\newcommand{\dcweight}{w_{H}}
\newcommand{\updateweight}{w_{u}}
\newcommand{\normal}{\mathbf{n}}
\renewcommand{\t}{\mathbf{t}}
\newcommand{\q}{\mathbf{q}}
\newcommand{\h}{\mathbf{h}}
\newcommand{\dvec}{\mathbf{d}}
\newcommand{\convergencethreshold}{\tau}

\usepackage{amsmath}
\DeclareMathOperator*{\argmin}{arg\,min}
\usepackage[ruled,linesnumbered,noend]{algorithm2e}

\DontPrintSemicolon

\SetKwFunction{SDFDualContouring}{sdf\_dual\_contouring}
\SetKwProg{Fn}{function}{}{}
\SetKw{KwBreak}{break}

\acmSubmissionID{654}

\makeatletter 
\newcommand{\layoutdetails}{%
\begin{tabular}{ll}
 \texttt{\textbackslash{textwidth}} & \printinunitsof{in}\prntlen{\textwidth} \\
\texttt{\textbackslash{linewidth}} & \printinunitsof{in}\prntlen{\linewidth} \\
Main text font &  \f@size pt \f@family \\
\sffamily \small Caption text font &  \sffamily \small \f@size pt \f@family \\
\end{tabular}%
}
\makeatother

\usepackage{xcolor,colortbl}
\definecolor{headercolor}{HTML}{b2df8a}
\definecolor{header2base}{HTML}{AED9E0}
\colorlet{header2color}{header2base!50}  %

\usepackage{graphicx}

\copyrightyear{2026}
\acmYear{2026}
\setcopyright{cc}
\setcctype{by-nc-nd}
\acmConference[SIGGRAPH Conference Papers '26]{Special Interest Group on Computer Graphics and Interactive Techniques Conference Conference Papers}{July 19--23, 2026}{Los Angeles, CA, USA}
\acmBooktitle{Special Interest Group on Computer Graphics and Interactive Techniques Conference Conference Papers (SIGGRAPH Conference Papers '26), July 19--23, 2026, Los Angeles, CA, USA}
\acmDOI{10.1145/3799902.3811116}
\acmISBN{979-8-4007-2554-8/2026/07}

\begin{document}

\title{Dual Contouring of Signed Distance Data}

\author{Xiana Carrera}
\affiliation{%
  \institution{Columbia University}
  \state{New York}
  \country{USA}
}
\email{x.carrera@columbia.edu}

\author{Ningna Wang}
\affiliation{%
  \institution{Columbia University}
  \state{New York}
  \country{USA}
}
\email{ningna.wang@columbia.edu}

\author{Christopher Batty}
\affiliation{%
  \institution{University of Waterloo}
  \city{Waterloo}
  \country{Canada}
}
\email{christopher.batty@uwaterloo.ca}

\author{Oded Stein}
\affiliation{%
  \institution{Technion}
  \city{Haifa}
  \country{Israel}
}

\affiliation{%
  \institution{University of Southern California}
  \city{Los Angeles}
  \state{CA}
  \country{USA}
}
\email{ostein@usc.edu}

\author{Silvia Sellán}
\affiliation{%
  \institution{Columbia University}
  \city{New York}
  \country{USA}
}
\email{silviasellan@cs.columbia.edu}

\renewcommand\shortauthors{Carrera et. al}

\begin{abstract}
  We propose an algorithm to reconstruct explicit polygonal meshes from discretely sampled Signed Distance Function (SDF) data, which is especially effective at recovering sharp features.
  Building on the traditional \emph{Dual Contouring of Hermite Data} method, we design and solve a quadratic optimization problem to decide the optimal placement of the mesh's vertices within each cell of a regular grid.
  Critically, this optimization relies solely on discretely sampled SDF data, without requiring arbitrary access to the function, gradient information, or training on large-scale datasets.
  Our method sets a new state of the art in surface reconstruction from SDFs at medium and high resolutions, and opens the door for applications in 3D modeling and design. 
\end{abstract}



\begin{teaserfigure}
  \includegraphics{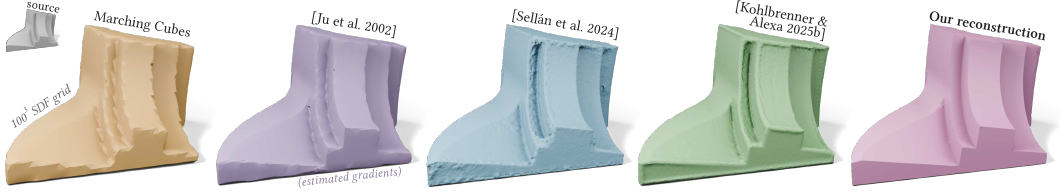}
  \caption{We propose an algorithm to reconstruct surfaces with sharp features from discrete signed distance data without the need for gradient information.}
  \label{fig:teaser}
\end{teaserfigure}

\maketitle

\section{Introduction}

\emph{Signed Distance Functions} (SDFs) measure the distance from any point in space to the boundary of a volume, using the sign to distinguish between the inside and the outside.
Their ability to represent shapes with complex sharp features to arbitrary precision and combine them efficiently using Boolean operations have made them popular in applications from computer graphics to engineering, industrial design and manufacturing.
In many of these, the SDF of a shape is sampled on a regular volumetric grid during the 3D modeling stage, the first in a larger pipeline with downstream tasks that require explicit surface representations (e.g., simulation).
Thus, reconstructing polygonal meshes from discrete SDF samples while preserving geometric features is a critically important task.

\begin{figure}[b]
\centering
\includegraphics{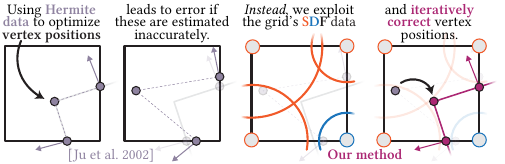}
\caption{Our work builds on the Dual Contouring method proposed by \citet{Ju2002}. Instead of relying on exact Hermite information, we estimate it first and progressively update it using discretely sampled SDF data.}
\label{fig:fig2}
\end{figure}

Unfortunately, existing reconstruction methods cannot accurately reconstruct a shape's sharp features from a finite set of SDF samples alone.
Recent work exploiting the global geometric information contained in every SDF sample has shown major improvements in reconstruction accuracy \cite{rfta,kohlbrenner2025polyhedral}; however, they rely on smoothness priors that soften a shape's sharp edges and corners (see \autoref{fig:teaser}).
Indeed, most prior work that succeeds at recovering sharp features does so by placing additional assumptions on the input: data-driven methods use neural networks trained over large datasets to develop powerful priors, while the traditional \emph{Dual Contouring of Hermite Data} \cite{Ju2002} assumes full knowledge of the surface's intersections with the grid edges and its local orientation at those points.

In this paper, we introduce a Dual Contouring algorithm designed to extract surfaces with sharp features from static sampled SDF data only.
Like the original work by \citet{Ju2002}, we assume that the samples are placed on the nodes of a regular grid, and design and solve a local quadratic energy minimization problem to select the optimal vertex placement within each cell.
Unlike the original Dual Contouring, our algorithm does not rely on exact surface normals: instead, it uses the SDF information itself to iteratively refine the mesh's vertex positions (see \autoref{fig:fig2}).

In contrast to a recent string of methods that improve reconstruction accuracy through complex optimizations that rely on the global information provided by SDF values away from the surface, our work uses a regular grid to build independent local optimizations that can be solved in parallel.
While our method \emph{can} exploit the global geometric information contained in SDF samples far from the zero level set, it is fundamentally agnostic to the amount of SDF data considered by each grid cell, which may be chosen to be higher or lower depending on computational constraints.

Through extensive qualitative and quantitative experiments, we demonstrate our method's superiority to prior work both in the reconstruction of sharp features and as a general reconstruction tool.
By phrasing reconstruction as a series of local optimizations and removing the need for intermediary representations like point clouds, our method avoids artificial smoothing and can recover sharp features even when these are not aligned with the background grid (see \autoref{fig:teaser}).
Our reconstruction strategy approaches the quality provided by the original dual contouring \cite{Ju2002}, but without the need for exact Hermite data.
Thus, our method sets a new state of the art in SDF reconstruction at medium and high resolutions, its benefits being most critical in applications in which precision and sharp feature recovery are particularly important, like manufacturing and industrial design.

\section{Related Work}

\begin{figure}
\centering
\includegraphics{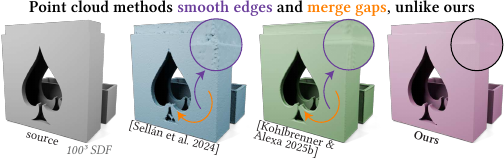}
\caption{Competing methods that apply Poisson reconstruction to an intermediate oriented point cloud suffer from over-smoothing. Our method preserves sharp edges and avoids merging close features.}
\label{fig:mind-the-gap}
\end{figure}

\begin{figure}
\includegraphics{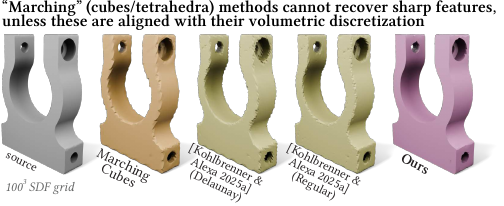}
\caption{Competing SDF reconstruction methods that rely on marching cubes or tetrahedra exhibit artifacts near sharp features, except when those features happen to align closely with the underlying elements of the volumetric mesh or grid. Our method exhibits no such bias.}
\label{fig:marching-methods-fail}
\end{figure}

SDFs can accurately represent topologically and geometrically complex solid shapes while enabling flexible editing and reasonably fast rendering \cite{Hart1996}.
Because of this, they have served as a useful tool in a myriad of scientific fields: from geology \cite{hayek2023diffuse} and medical imaging \cite{esposito2025vesselsdf} to computational fluid dynamics \cite{sethian2003level}, robotic path planning \cite{Liu2022} and additive manufacturing \cite{Brunton2021}.
In computer graphics, SDFs have been used to represent the free surface of a fluid \cite{foster2001practical} or detect collisions between objects \cite{fuhrmann2003distance}; more recently, their representational power has been exploited in geometric deep learning and generative 3D modeling \cite{Park2019,takikawa2021nglod,Sharp2022,wang2023neural,coiffier20241,Marschner2023}.
They have even been adopted as the geometric representation of choice by computational design companies like \citet{ntop} and \citet{metafold}.

Despite the benefits of SDFs, they are ill-suited for many downstream tasks: from finite element simulation to texture mapping and real-time rendering, applications often rely on converting discretely sampled SDF solids into explicit representations like polygonal meshes on which these tasks can be efficiently performed.
Thus, reconstructing meshes from SDFs has long been a critical research question in the geometry processing community.
In the next two sections, we review the most relevant existing SDF isosurfacing algorithms, starting from those that (like ours) rely only on discretely sampled SDF data as input and ending with those that place additional assumptions on the input.

\subsection{Reconstructing discretely sampled SDFs}

A vast amount of prior research has been dedicated to the more general problem of reconstructing meshes from arbitrary implicit representations: to borrow the categorization proposed by \citet{deAraujo2015}, these methods may opt for region-growing \cite{hilton1996marching}, inflation/shrinkwrap schemes \cite{stander1997}, or, more commonly, spatial discretizations like regular grids (e.g., \emph{Marching Cubes} \cite{Lorensen1987}).
As shown by \citet{rfts}, these methods are by and large designed for general implicit representations; when applied to SDFs, they are not as accurate as purpose-made algorithms that exploit the distance information contained in each sample.
We focus our discussion on the latter and refer the interested reader to the survey by \citet{deAraujo2015} for a comprehensive review of implicit isosurfacing.

\citet{rfts} proposed interpreting global signed distance data as imposing a series of tangency constraints on the reconstructed surface, which they enforce weakly by minimizing an \emph{SDF energy} through a mesh-based geometric gradient flow. This gradient flow includes a local remeshing step that smooths a surface's sharp features; more critically, it encounters singularities and fails to converge if the shape's topology is complex or not known beforehand. 

To avoid singularities and support arbitrarily complex topologies, \citet{rfta} proposed using SDF data to construct an intermediate point cloud representation, which can then be passed to an off-the-shelf point-based reconstruction algorithm \cite{Kazhdan2006,Kazhdan2013}.
The crux of the method is in the positioning of the points: while \citet{rfta} propose a complex optimization involving rasterization and iterative refinement, the follow-up work by \citet{kohlbrenner2025polyhedral} uses Lie geometry to produce a valid, more regular point cloud.
Regardless of how the point cloud is computed, these methods inherit the smoothness prior of the specific point cloud reconstruction algorithm, which causes smoothed corners and merged gaps (see Figures~\ref{fig:teaser} and \ref{fig:mind-the-gap}).

Exploiting the same insight, \citet{kohlbrenner2025isosurface} use SDF data to construct a volumetric regular tetrahedralization, from which an isosurface can be extracted using marching tetrahedra.
More simple and computationally efficient than point and flow-based algorithms, it outperforms other spatial subdivision methods like Marching Cubes by grouping resolving power closer to a shape's important features. However, the edges of the regular tetrahedralization are not guaranteed to align with a shape's sharp features, leading to smoothed and chamfered corners (see \autoref{fig:marching-methods-fail}).

Our method builds on the observations underpinning recent methods for reconstruction from discretely sampled SDFs, addressing some of their main limitations. We extract the mesh's topology from a regular grid, avoiding singularities and unstructured remeshing. We do not rely on any intermediate representation, thus removing the need for smoothness priors (see Figure 3) and superlinear complexity. Finally, we use a dual spatial discretization, recovering sharp features irrespective of their grid alignment (see \autoref{fig:marching-methods-fail}).

\subsection{Dual Contouring of Hermite Data}

\begin{figure}
\centering
\includegraphics{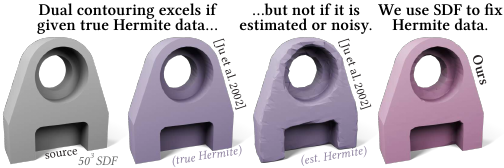}
\caption{The outstanding reconstruction quality of classic Dual Contouring depends on having \emph{exact} Hermite data as input (center left). If approximate (finite difference) estimates are used, quality suffers noticeably (center right). Our method uses the SDF data to optimize for accurate Hermite data, nearly recovering the output quality achieved with exact data (far right).}
\label{fig:dc-fails-if-Hermite-bad}
\end{figure}

Mathematically, our method is most related to and greatly inspired by the \emph{Dual Contouring of Hermite Data} algorithm proposed by \citet{Ju2002}.
When combined with follow-up work improving stability and geometric properties \cite{schaefer2002dual,ju2006intersection,schaefer2007manifold,trettner2020fast}, it produces impressively accurate reconstructions with close-to-perfectly recovered sharp features (see \autoref{fig:dc-fails-if-Hermite-bad}, left).

Its effectiveness is in part thanks to the format of its input: it requires exact knowledge of both the positions at which the grid edges intersect the unknown surface and the local normal directions there (from here onwards, we refer to such point-normal pairs as \emph{Hermite data}).
While this data can be extracted from a signed distance function if one has the ability to query the SDF at arbitrary spatial positions (e.g., through bisection along each edge combined with finite difference gradient estimation), it can only be loosely approximated if the SDF's values are only known on a discrete set of locations.
If this estimation is inaccurate (as is often the case near an object's sharp features), the method loses its ability to reconstruct sharp features (see \autoref{fig:dc-fails-if-Hermite-bad}, center-right). 

\begin{figure}
\centering
\includegraphics{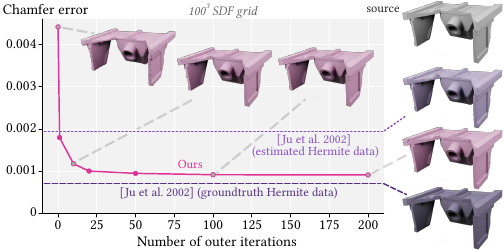}
\caption{Our method progressively corrects an estimated set of Hermite data until our reconstruction approaches the quality produced by the original Dual Contouring with groundtruth Hermite data.}
\label{fig:ablation-outer-iter}
\end{figure}

Like \citet{Ju2002}, we propose a contouring strategy that optimizes vertex placements within each cell using a quadratic error function (QEF). However, we do so without requiring exact Hermite data or access to the SDF function, and instead rely only on a discrete set of SDF samples. We exploit the information contained in these samples to iteratively correct an incorrect initial set of Hermite data, eventually recovering sharp features with nearly the same quality achieved by \citet{Ju2002}, but without placing restrictive assumptions on the input (see \autoref{fig:ablation-outer-iter}).

\subsection{Neural SDF reconstruction}
Neural techniques have had significant recent influence on isosurface extraction of SDF data. It is helpful to disambiguate between two different tasks that are nonetheless often referred to in the literature by the same name. Some methods are dedicated to meshing \emph{continuous} SDFs  represented (perhaps approximately) by neural networks \cite{xie2022neural,hwang2024occupancy,liu2025direct,remelli2020meshsdf,shen2023flexible,binninger2025tetweave,lei2020analytic,stippel2025marching}, while others concern themselves with learning data-driven priors to mesh \emph{discretely sampled} SDFs.

\begin{figure*}
\centering
\includegraphics{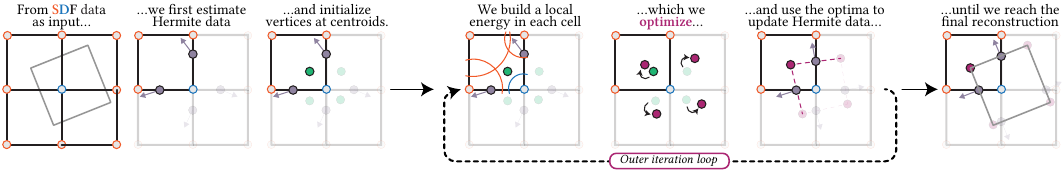}
\caption{An overview of the outer optimization loop of our method. After estimating Hermite data at each interesting edge, an initial configuration is calculated and progressively refined through local per-cell optimizations and global updates to the Hermite data.}
\label{fig:method-summary}
\end{figure*}

The latter category of methods is most relevant to our setting. Neural Marching Cubes \cite{Chen2021} and Neural Dual Contouring \cite{Chen2022} update their classic counterparts by learning the placement of vertices from datasets of meshes, and using wider stencils of SDF data for inference compared to the classic methods. In Neural Dual Contouring, the role of the QEF is entirely replaced by a learned network, obviating the need for Hermite input. 
Rather than replacing the QEF, the PoNQ method \cite{maruani2024ponq} proposes a point-based surface representation augmented with QEF data. They learn to generate the PoNQ data using a network that accepts as input the entire $N^3$ SDF grid. Surface reconstruction is then performed using a Delaunay-based approach that tetrahedralizes the optimal QEF points (analogous to Dual Contouring vertices) and tags each tetrahedron as interior or exterior with a graph-cut strategy, ensuring a watertight and non-self-intersecting surface. PoNQ's predecessor VoroMesh \cite{maruani2023voromesh} uses a related Voronoi-based strategy, but without QEF data. As with any learning-based method, these approaches require significant training time over large data sets, and their results will be dependent on the characteristics of the training data.

\subsection{Quadratic Error Functions in Computer Graphics}

To produce accurate surface reconstructions with sharp feature edges, we design a tailor-made quadratic optimization problem that is solved to obtain the optimal vertex placement inside each grid cell.
This puts our method, like the original by \citet{Ju2002}, in the company of a vast array of geometry processing algorithms. Indeed, quadratic error functions have been used for tasks from mesh simplification \cite{garland1998simplifying} and approximation \cite{thiery2013sphere} to surface fairing \cite{legrand2019filtered} and building simulation cages \cite{deng2011automatic}.

More relevant to our work, quadric error metrics have been used to reconstruct surfaces from signed \cite{liu2025direct} and unsigned \cite{zhang2023surface} distance data.
While these methods produce impressive results, even in the presence of noise, they do so by exploiting the ability to sample the distance function at arbitrary spatial positions, enabling them to build the optimal quadratic error function.
Our work considers a related but critically different task: namely, that of reconstructing surfaces with sharp features from discrete SDF samples \emph{alone}, without access to more data or the ability to query the SDF function at additional locations.

\section{Method}

\begin{figure}
\centering
\includegraphics{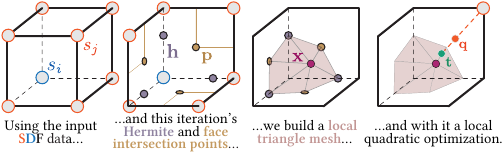}
\caption{In each iteration, we use the Hermite (edge) points and face intersection points to build a local mesh inside each cell, whose interior vertex we optimize to align with SDF values assigned to the cell.}
\label{fig:local-mesh}
\end{figure}

Our input is a uniform, regular grid equipped with the values $\sdf_1,\dots,\sdf_n\in\R$ measuring the signed distance from an unknown surface $\Omega$ to each of the grid's vertices $\cellcorner_1,\dots,\cellcorner_n \in \R^3$. In this section, we will introduce an iterative algorithm to compute a quad mesh $\mesh = (\cellverts,\allquads)$ that approximates $\Omega$.

As is customary in grid-based SDF reconstruction algorithms (e.g., \cite{Lorensen1987,Ju2002}), we begin by identifying the grid's \emph{interesting} edges, i.e., those whose vertices differ in sign and are thus known to intersect the true surface $\Omega$. 
We then tag all cells containing any such edge as \emph{interesting cells}, the set of which we will denote $\intCells = \cell_1,\dots,\cell_m$. 

Following the contouring strategy proposed by \citet{Ju2002}, we will now phrase the reconstruction problem as that of optimizing the placement of a single vertex $\cellvert_i$ in each interesting cell $\cell_i$.
Once the ideal vertex positions $\cellvert_1,\dots,\cellvert_m$ have been found, the four vertices corresponding to the four cells containing each interesting edge can be joined together to form our mesh's quads $\allquads$.

\subsection{Initialization}

To compute an initial guess for each cell's optimal vertex placement, we begin by looping over all interesting edges $\edge_i = [\cellcorner_{ia},\cellcorner_{ib}]$. Given the difference in sign between the two edge's SDF values, linear interpolation allows us to estimate the point along the edge that intersects with the surface $\Omega$,
\begin{equation}\label{equ:Hermitepointest}
  \Hermitepoint_i^0 = (1-t)\cellcorner_{ia} + t\cellcorner_{ib}\,,\quad \text{where}\quad t = |\sdf_{ia}|/\left(|\sdf_{ia}| - |\sdf_{ib}|\right)\, .
\end{equation}

\begin{figure*}
\centering
\includegraphics{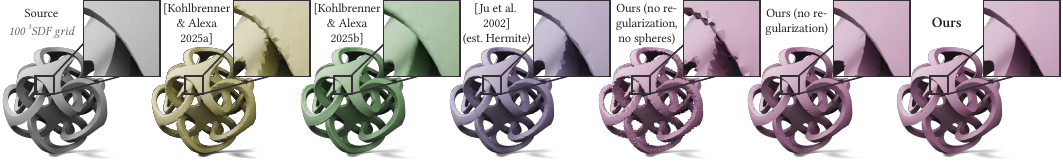}
\caption{To guide convergence in a highly non-convex setting, our algorithm incorporates two regularization terms in our linearized inner loop, one based on the estimated Hermite normals and the other a standard $L_2$ regularizer.}
\label{fig:energy-terms}
\end{figure*}

We then differentiate the trilinear interpolant to construct gradients over each interesting cell, $\gradtrilinear(\cell_1,\mathbf{x}), \dots, \gradtrilinear(\cell_m,\mathbf{x})$. We evaluate the gradient at the Hermite point of edge $\edge_i$ for each cell containing the edge and average them to estimate the Hermite \emph{normal}:
\begin{equation}\label{equ:Hermitenormalest}
  \Hermitenormal_i^0 = \scriptstyle \frac{\sum_{\edge_i\in\cell_j} \gradtrilinear(\cell_j, \Hermitepoint_i^0)}{\left\|\sum_{\edge_i \in \cell_j} \gradtrilinear(\cell_j,\Hermitepoint_i^0)\right\|}.
\end{equation}

With this estimated information, one could directly apply the original Dual Contouring algorithm to obtain a possible reconstruction.
In practice, we note that this strategy (which we refer to as \emph{\cite{Ju2002} (estimated Hermite data)} throughout the paper) fails to reconstruct the shape's features, as the estimates $\{( \Hermitepoint_i^0, \Hermitenormal_i^0 )\}$ are least accurate near sharp gradient discontinuities (see \autoref{fig:dc-fails-if-Hermite-bad}).
Nonetheless, one may imagine that the vertex positions produced by their algorithm can serve as a helpful initialization for our method;
unfortunately, this is not the case, as they often consist of irregular shapes with self-intersections that in practice can trap our iterative reconstruction in undesirable local minima.

Instead, we opt for a more regular initialization strategy. Once this first estimate for each edge's Hermite information has been calculated, we compute each interesting cell's \emph{centroid} by averaging all Hermite points contained in its edges:
\begin{equation}
  \centroid_i^0 =  \frac{\sum_{\Hermitepoint_j^0 \in \cell_i} \Hermitepoint_j^0}{|\{\Hermitepoint_j^0 \in \cell_i\}|}.
\end{equation}
We then initialize each cell's vertex to its centroid: $\cellvert_i^0 = \centroid_i^0$.
In the rest of this section, we will proceed iteratively to improve the positioning of this vertex. In particular, given the input SDF values and a prior iteration's cell vertices $\cellvert_i^k$, we will construct an energy minimization problem whose solution provides an updated position $\cellvert_i^{k+1}$.
We refer to the repetition of this process as our algorithm's \emph{outer} iterative loop, which we discuss in the next section.

\subsection{Outer iterative loop}

Given a prior iteration's vertex positions $\cellvert_i^k$, we connect the four cell vertices corresponding to each interesting edge to form a quad mesh, which we refer to as this iteration's \emph{global} mesh $\mesh^k$.
Inspired by a long line of work from the geometry processing community proposing local-global optimization strategies \cite{arap}, our goal now is to use the information contained in this mesh and the input SDF data to construct a \emph{local} energy minimization problem that can be solved in parallel to obtain each cell's new vertex position.

Making each cell's optimization problem depend on all the input SDF data would be computationally prohibitive.
Because of this, we start by arbitrarily triangulating the global mesh and computing the closest point on the mesh to each of the grid nodes $\cellcorner_j$.
We identify which of the grid cells contains this closest point, and \emph{assign} the tuple $(\cellcorner_j,\sdf_j)$ to said cell.
For robustness to outliers, we do not assign any sphere whose distance to the closest point is larger than the sum of the cell diagonal and $|\sdf_i|$.

We will now quantify how much a given choice of cell vertex aligns with the SDF values $(\cellcorner_j,\sdf_j)$ assigned to the cell. To achieve this, we will use the current iteration's vertex $\cellvert_i$ and Hermite data $\Hermitepoint^k$ and construct a \emph{local} triangle mesh that approximates the reconstruction's geometry at $\cell_i$ without introducing explicit dependencies between cells (thus maintaining locality and parallelization).

By construction, the global mesh's edges will cross the planes of the interesting cell's faces: we compute and store these intersections or \emph{face intersection points} $\fip^k$.
Each $\fip^k$ is spawned by a specific quad in the mesh, and each quad corresponds to one of the grid's interesting edges. That edge has itself an associated Hermite point $\Hermitepoint^k$, which we say is $\fip^k$'s \emph{relevant Hermite point}.
A cell's local triangle mesh $\mesh_i^k$ can then be constructed by joining together every face intersection point, its associated Hermite point, and the cell's vertex $\cellvert$ (see \autoref{fig:local-mesh}, left), inspired by the approach used by Occupancy-Based Dual Contouring \cite{hwang2024occupancy}.

The extent to which $\mesh_i^k$ agrees with the cell's assigned SDF data can be measured in terms of the local \emph{distance energy}
\begin{equation}\label{equ:distanceenergy}
\sdfenergy^{k,i}(\cellvert) = \sum\limits_{(\cellcorner_j,\sdf_j) \in \assigned(\cell_i)} \left( |\sdf_j| - \distance\left(\cellcorner_j,\mesh_i^k(\cellvert)\right)\right)^2
\end{equation}
where $\assigned (\cell_i)$ is the set of $(\cellcorner_j,\sdf_j)$ tuples assigned to the cell $\cell_i$, and $\distance$ measures the Euclidean distance from a point to a triangle mesh.

This energy is inspired by the \emph{SDF energy} proposed by \citet{rfts}; however, because it is applied to an open local mesh as opposed to the full reconstruction, it simply ignores the distance’s sign. Because our method uses a regular grid discretization that only locally updates each cell’s vertex position, our output topologically separates the positive SDF data from the negative one by construction, removing the added complexity of a signed energy treatment.
By contrast, the method of \citet{rfts} relies on a global geometric flow that optimizes an entire mesh starting from a large, encompassing initial guess (e.g., a sphere); it must therefore account for the distances’ signs to ensure SDF samples  $(\cellcorner_j,\sdf_j)$ end up on the correct side of the surface. This need poses an additional challenge in their energy minimization, requiring a purpose-made metric of sign agreement based on the surface’s normals.

\begin{figure}
\centering
\includegraphics{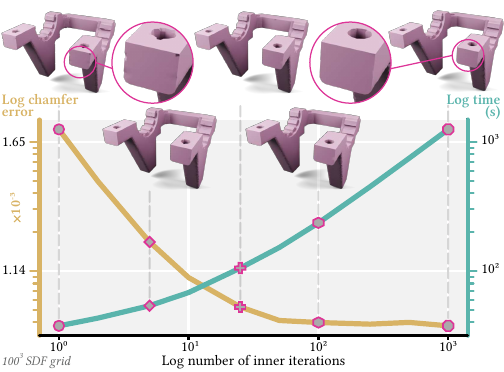}
\caption{Increasing the number of inner iterations progressively reduces the reconstruction error, eliminating local artifacts and improving feature fidelity until accuracy gains saturate, while runtime grows roughly linearly.}
\label{fig:ablation-inner-iter}
\end{figure}

Directly minimizing our distance energy will produce highly irregular solutions (see \autoref{fig:energy-terms}): intuitively, this is because the union of the local meshes $\mesh_i^k$ is only a valid approximation of the global mesh for reasonably small displacements in $\cellvert$. In particular, minimizing \autoref{equ:distanceenergy} separately for all the vertices in a single quad may lead to optimal local meshes that are artificially bent in ways that are impossible to recreate with a single quadrilateral.
Thus, we also include the \emph{Hermite energy} (c.f. \citet{Ju2002}):
\begin{equation}\label{equ:dcenergy}
  \dcenergy^{k,i}(\cellvert) = \sum\limits_{\edge_j \in \cell_i} \left( (\cellvert - \Hermitepoint_j^k) \cdot \Hermitenormal_j^k \right)^2.
\end{equation}
In our setting, this energy plays a dual role. First, given that each Hermite tuple $\{(\Hermitepoint_j^k, \Hermitenormal_j^k)\}$ is shared between all cells that will form a quad together, it encourages a consistent normal orientation, ensuring that the distance energy is not satisfied through artificial bending in the local mesh.
Crucially, this reduces the importance of the choice of triangulation in later steps of our method (L11 in Algorithm \ref{alg:pseudocode}).
Second, since our outer iterations are also designed to progressively improve the Hermite data itself, this energy helps mimic the sharp edge recovery behavior that \citet{Ju2002} achieved through exact Hermite data.
The combination of Equations~\ref{equ:distanceenergy} and \ref{equ:dcenergy} lead to our total \emph{local energy}
\begin{equation}
\energy^{k,i} (\cellvert) =  \sdfenergy^{k,i}(\cellvert) + \dcweight^2 \dcenergy^{k,i}(\cellvert)\,,
\end{equation}
where $\dcweight$ is a weight balancing the effect of the two terms (see \autoref{fig:ablation-dc-weight} for an ablation study on the effect of this parameter).
In \autoref{sec:innerloop}, we will show how we turn this energy into an iterative quadratic problem (our \emph{inner iterative loop}), enabling us to solve it in parallel for all cells during each outer iteration.

\begin{figure}
\centering
\includegraphics{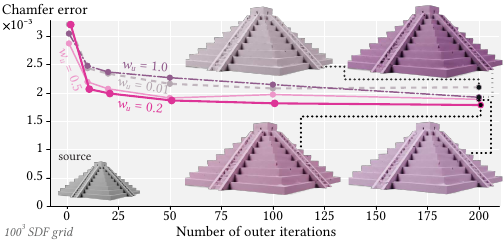}
\caption{Increasing the update weight $\updateweight$ may not always improve accuracy when the number of outer iterations is low, but choosing very conservative values can make precision saturate too soon. We find intermediate values in the range 0.2-0.8 to generally yield the best results.}
\label{fig:ablation-uw}
\end{figure}

Once this energy has been optimized for every cell to obtain the optimal next vertex positions $\cellvert_i^{k+1}$, our next and final step is to update the Hermite information in each interesting edge.
For every four cells sharing a Hermite edge, we use Principal Component Analysis on the associated four vertices to obtain their best-fit plane, normal $\normal$, and intersection with the edge $\mathbf{y}$. Then, we  update the Hermite information at that edge using linear interpolation:
\begin{equation}\label{equ:Hermiteupdate}
  \Hermitepoint_j^{k+1} = \Hermitepoint_j^{k} + \updateweight (\mathbf{y} - \Hermitepoint_j^{k})\,, \quad \Hermitenormal_j^{k+1} = \frac{\normal + \updateweight \Hermitenormal_j^{k}}{\|\normal + \updateweight \Hermitenormal_j^{k}\|}\,,
\end{equation}
where our \emph{update weight} $\updateweight \in [0,1]$ provides a tradeoff between stability and speed of convergence (see \autoref{fig:ablation-uw} for an ablation). In the last iteration, we join together the final cell vertices $\cellverts = \cellvert_1,\dots, \cellvert_m$ to form our output quad mesh $\mesh = (\cellverts,\allquads)$.

\subsection{Inner iterative loop}\label{sec:innerloop}
We now describe how we minimize the local energy $\energy^k$ for each cell in each iteration $k$ of our outer loop (for clarity, we omit the superscript $k$ from here onwards), a highly non-convex optimization problem difficult to optimize for directly with any guarantees.
Like \citet{rfts}, we begin by interpreting each pair of SDF data $(\cellcorner_j,\sdf_j)$ as a sphere centered at $\cellcorner_j$ with radius $|\sdf_j|$ to which the local mesh must be tangent. Letting $\t_j$ be the closest point to $\cellcorner_j$ on the local mesh $\mesh_i$ and $\q_j$ the closest point to $\t_j$ on the surface of the sphere, one can linearize each term in the distance energy as
\begin{equation}\label{equ:taylorsversion}
\left( |\sdf_j| - \distance\left(\cellcorner_j,\mesh_i\right)\right)^2 \approx \left\| \t_j - \q_j\right\|^2\,.
\end{equation}
This linearization is sufficient in global optimization settings, in which the interaction between different mesh vertices and even remeshing can aid one in escaping local minima \cite{rfts}.

\begin{figure}
\centering
\includegraphics{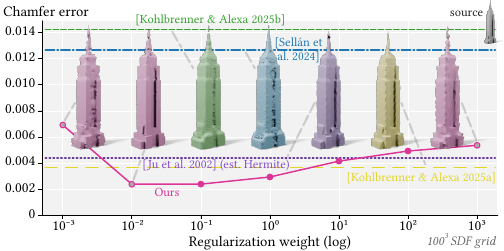}
\caption{\emph{Start spreading the $\mu$'s}: our algorithm benefits from inter-iteration regularization; in practice, we recommend values of $\mu\in[10^{-2},10^{-1}]$.}
\label{fig:ablation-mu}
\end{figure}

\begin{figure*}[]
\centering
\includegraphics{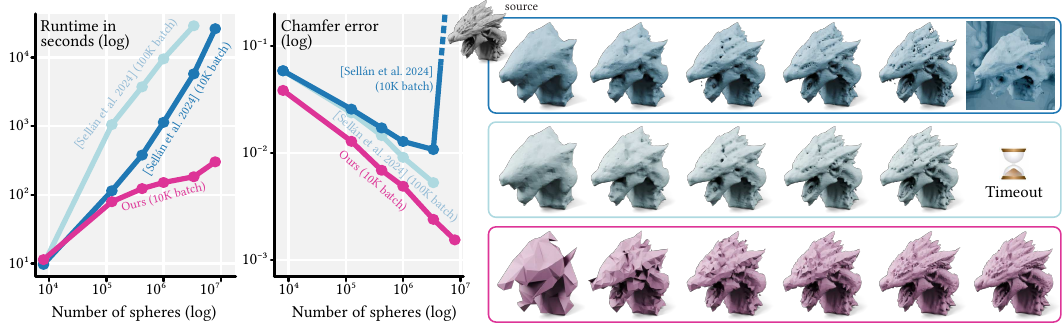}
\caption{We compare our method (pink) against \citet{rfta} using their default batch size (10K, blue) for the number of spheres being considered, and an increased batch size (100K, cyan), which rapidly makes computational costs intractable. Not using a high enough batch size with their method can cause parts of the surface to be generated far away from the ground truth, fully enclosing the reconstruction (top row, rightmost).}
\label{fig:ours-vs-rfta-error-and-runtime}
\end{figure*}

\begin{figure}
\centering
\includegraphics{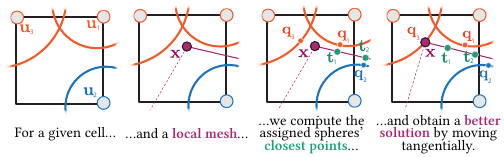}
\caption{An example in which allowing tangential displacements and penalizing only the radial distance toward the spheres' centers yields a more accurate vertex position.}
\label{fig:projection}
\end{figure}

However, it is not enough in our local setting, in which a cell's vertex may be prevented from approaching an unsatisfied sphere because sliding tangentially with respect to others is penalized. Figure \ref{fig:projection} gives an example in which the cell's vertex $\cellvert$ cannot move towards $\q_3$ because it would cause $\t_1$ and $\t_2$ to slide away from their respective $\q_i$.
To avoid this, we will only measure distance in the direction towards the sphere's center, and not tangential to it. Letting $\dvec_j$ be the (normalized) vector from $\cellcorner_j$ to $\q_j$, the distance energy becomes
\begin{equation}
\left( |\sdf_j| - \distance(\cellcorner_j,\mesh_i)\right)^2 \approx \left( (\t_j - \q_j)\cdot \dvec_j \right)^2\,.
\end{equation}
Formally, this amounts to a first-order Taylor expansion of the distance in \autoref{equ:taylorsversion}.
Writing $\t_j$ in terms of its barycentric coordinates in the local triangle mesh allows us to write the right hand side of the above approximation as a quadratic energy in $\cellvert$:
\begin{equation}
  \tilde{\sdfenergy}(\cellvert) = \|A_d\mathbf{x} - b_d\|^2\,
\end{equation}
where $A$ and $b$ concatenate the contribution of each relevant tuple $(\cellcorner_j,\sdf_j)$. To this linearized distance energy, we add the already-quadratic Hermite energy from \autoref{equ:dcenergy}. We will minimize them together, starting from the previous iteration's cell vertex $\cellvert^{k} = \cellvert^{k,0}$ and iteratively updating the positions $\t_j,\q_j$ using each subsequent minimizer. Finally, we include a regularization term to compensate for the aggressiveness of our linearization:
\begin{equation}\label{equ:xkrp1}
  \cellvert^{k,r+1} = \argmin\limits_{\cellvert} \tilde{\sdfenergy}^{k,r}(\cellvert) + \dcweight^2\Hermiteenergy^{k}(\cellvert) + \mu\|\cellvert - \cellvert^{k,r}\|^2.
\end{equation}

\begin{figure}
\centering
\includegraphics{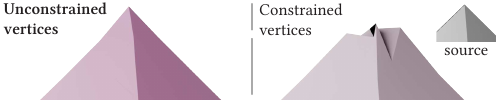}
\caption{Allowing vertices to move outside their corresponding cells helps reconstruct particularly sharp features.}
\label{fig:contrain-vertices-to-cell}
\end{figure}

Whenever the distance between one $\cellvert^{k,r}$ and the next $\cellvert^{k,r+1}$ is lower than a chosen numerical threshold $\convergencethreshold$, or when a maximum number of iterations is reached, the algorithm returns to the outer loop with an updated $\cellvert^{k}$.
While we experimented with constraining $\cellvert^{k,r+1}$ to always be inside the cell, we found that allowing it (and the face intersection points $\fip$) to escape the cell helped with sharp feature recovery (see Figure  \ref{fig:contrain-vertices-to-cell}).

\section{Experiments \& Results}\label{sec:results}

\paragraph{Implementation.} We developed our method in C++ and Python, using libigl \cite{libigl} and Gpytoolbox \cite{gpytoolbox} for common geometric processing subroutines (e.g., AABB trees for point-to-mesh distances) and OpenMP to perform our vertex optimizations in parallel for each cell.
We report runtimes on a Macbook Pro laptop with an M4 Pro chip and 48 GB of memory, and render our results using BlenderToolbox \cite{blendertoolbox}.
Our comparisons to the methods of \citet{rfts}, \citet{rfta}, \citet{kohlbrenner2025polyhedral} and \citet{kohlbrenner2025isosurface} rely on the authors' official implementations.
Our method's code is publicly available online\footnote{\url{https://gatc.cs.columbia.edu/projects/dual-contouring-of-signed-distance-data.html}}, and we provide pseudocode in the \textbf{Supplementary Material}.

\begin{figure}
\centering
\includegraphics[width=3.36in]{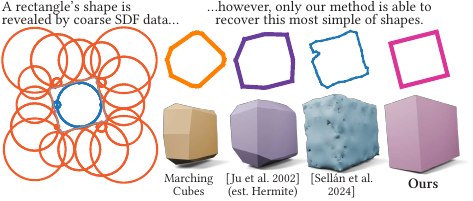}
\caption{Even at very low resolution ($5^3$), the sphere visualization of this SDF makes the shape of a source rectangle apparent (left). Our method easily recovers the rectangle (top row), and its counterpart 3D cuboid (bottom row), while other methods fail.}
\label{fig:cubes-experiment}
\end{figure}

\begin{figure}
\centering
\includegraphics{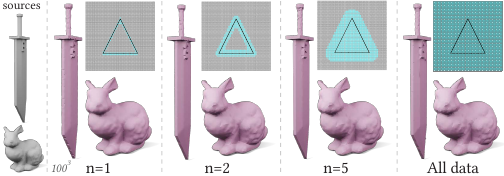}
\caption{Our algorithm successfully reconstructs the input even when using a limited amount of information. Here, we only consider the SDF data available at grid vertices whose distance from the ground truth is smaller than the length of the diagonal of a cell grid scaled by $n$.}
\label{fig:narrow-band}
\end{figure}

\paragraph{Experiments}
Our algorithm's runtime is dominated by the quadratic optimization steps in our inner loop, whose computational cost scales linearly with the size of the input data.
Fortunately, our algorithm is agnostic to the structure of this input: at low resolutions, one may want to exploit all the information contained in a regular grid (e.g., in \autoref{fig:cubes-experiment}); at higher ones with more redundancy, one may wish to use only a narrow band around the zero level set (see \autoref{fig:narrow-band}).
If one wishes to exploit all the data in the input even in higher resolution settings, we can randomly batch the input data considered in each outer iteration: in Figures~\ref{fig:ours-vs-rfta-error-and-runtime} and \ref{fig:asymptotic-complexity}, we show that this strategy improves asymptotic complexity with minimal effect on accuracy. In all experiments, we set a maximum batch size of $200{,}000$.
As expected, our method's complexity also scales linearly with the number of inner and outer optimization iterations performed: in Figures \ref{fig:ablation-outer-iter} and \ref{fig:ablation-inner-iter}, we show that diminishing returns are observed around $100$ inner and outer maximum iterations, which we set as our default values.

Beyond its runtime, several other parameters affect our method's performance. As shown in \autoref{fig:ablation-dc-weight}, it is most sensitive to the choice of $\dcweight$, which we fix to $0.02$ in all our experiments.
The two other major parameters of our algorithm are the update weight $\updateweight$ and the regularization $\mu$: in \autoref{fig:ablation-uw} and \autoref{fig:ablation-mu}, we study the effect of these parameters, whose defaults we set to $0.2$ and $0.1$ respectively.
Every result in our paper includes a randomly generated rotation, to control for the effect of axis alignment.

\paragraph{Comparisons \& Applications}
We carry out several quantitative and qualitative experiments to test the performance of our method, beginning with the one that inspired this research. In \autoref{fig:cubes-experiment}, we attempt a deceptively simple task: to reconstruct a (2D or 3D) box from a coarse, discrete set of SDF samples alone, without differential or Hermite information.
Surprisingly, no previously existing algorithm managed to successfully produce sharp corners, regardless of size and orientation (unless it perfectly aligns with the background grid).
We used this example during the development of our algorithm; in its final form, it recovers the perfectly sharp box.

To evaluate our method in a setting closer to its real-world applications, we considered the randomly selected subset of the ABC dataset \cite{Koch_2019_CVPR} released by \citet{xu2024cwf}, and used it to produce SDF samples at resolutions $50^3,100^3,150^3$ and $200^3$. We pass these samples as input to our method as well as Marching Cubes \cite{Lorensen1987} and the methods of \citet{Ju2002} (with estimated Hermite data), \citet{rfts}, \citet{rfta}, \citet{kohlbrenner2025isosurface}, and \citet{kohlbrenner2025polyhedral}.
As proposed by these authors, we quantify Hausdorff and chamfer error between groundtruth and output, as well as the SDF energy introduced by \citet{rfts}. To evaluate accuracy in sharp feature recovery, we also use the \emph{edge} chamfer error proposed by \citet{Chen2021}.

We present our results in \autoref{tab:quant_results} (\textbf{Supplemental}) and \autoref{fig:large-scale-test}, which reveal our method to have the highest accuracy across resolutions and almost all metrics.
Surprisingly, the second best scores are often achieved by the original Dual Contouring method with estimated Hermite data, revealing that the high reported accuracy of prior point-cloud based work (e.g., \cite{rfta,kohlbrenner2025polyhedral}), which was tested mostly on smooth organic shapes, may not carry over to cleaner CAD geometry with sharp features.
In \autoref{fig:mind-the-gap}, we show a qualitative result illustrating the failure cases of these algorithms.
At the highest resolutions, \emph{Reach for the Arcs}' \cite{rfta} runtime made it impracticable, and the method proposed by \citet{kohlbrenner2025polyhedral} encountered crashes for a large subset of shapes. Therefore, at these resolutions, we compare only to the method by \citet{kohlbrenner2025isosurface}, which our algorithm outperforms. In \autoref{fig:marching-methods-fail}, we qualitatively compare to this method, which fails to recover precise sharp features.

Our method's impressive performance on this dataset reveals that it is most accurate (in relation to other methods) in medium-to-high resolution settings, and on the broadly smooth shapes with sharp gradient discontinuities that dominate the ABC dataset.
To test this intuition, we evaluate our method on the ten artist-designed smooth geometries used as a baseline by \citet{rfta}, and report the results in \autoref{tab:quant_results_10shapes}.
The results of this specific dataset without sharp features are more mixed, varying based on resolution and metric. In rough terms, we observe our algorithm to perform on par with the best of the previously existing ones.

\begin{figure}
\centering
\includegraphics{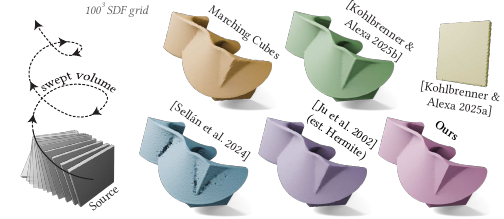}
\caption{Sweeping a rigid rectangular plate along a swirling trajectory yields a complex shape with numerous sharp features. Our method produces the best reconstructed mesh compared to competing schemes.}
\label{fig:swept-volumes}
\end{figure}

\begin{figure}

\centering
\includegraphics{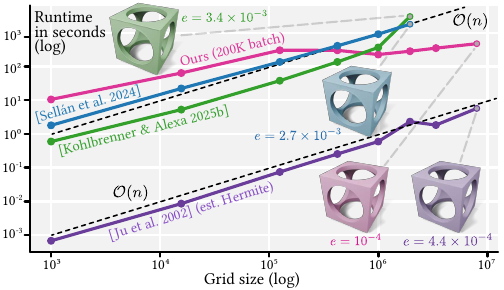}
\caption{While prior point-based methods scale superlinearly or linearly with the size of the SDF data (even the work by \citet{rfta}, which includes constant-size batching), our random asymptotic complexity is bounded by the size of the batch.}
\label{fig:asymptotic-complexity}
\end{figure}

Our method will be most impactful in applications in which one wishes to reconstruct shapes with sharp features at a high resolution and it is impossible or exceedingly costly to estimate exact gradient information or densely sample the SDF.
In \autoref{fig:swept-volumes}, we show one such example, in which each SDF sample is painstakingly computed by taking the minimum over tens of thousands of positions along an object's trajectory.
In this context, our method recovers sharp edges without the need for binary search of gradient computation, considerably speeding existing swept volume computation frameworks \cite{sellan2021swept}.
For this result, we follow the lead of \citet{rfts} and do not impose tangency constraints for the \emph{pseudoSDF} samples on the inside of the swept volume \cite{Marschner2023}.

\paragraph{Limitations \& Conclusion}
We have introduced an algorithm to reconstruct polygonal meshes from discretely sampled signed distance functions.
Building on the original work by \citet{Ju2002}, we proposed a Dual Contouring strategy that begins by estimating Hermite information on a grid's edges and progressively corrects it by exploiting the information contained in the SDF data.
Unlike all prior work on this task, our algorithm is able to faithfully reconstruct a shape's sharp edges through the SDF information, without the need for gradient information or data-driven priors.

Our method inherits some limitations from the original Dual Contouring. For example, we recover sharp features by optimizing the position of a vertex associated with each of the grid's cells.
Often, the most accurate reconstruction relies on this vertex escaping outside the bounds of the cell, which can cause self-intersections and face flips (see Figure \ref{fig:ablation-dc-weight}). Similarly, as we study in \autoref{tab:quant_results_noise}, our method inherits the original Dual Contouring's sensitivity to noise.

To obtain accurate reconstructions efficiently, our method relies on a heavily non-convex energy, whose minimizer is only approximated numerically through a local-global iterative optimization. In challenging cases, this approximation can yield isolated dents along feature curves (see Figure \ref{fig:ablation-dc-weight}). We conjecture that they might be alleviated through a more complex global mesh update that enforces sparsified regularity in each iteration's Hermite data.

After optimizing the vertex corresponding to each cell, we connect vertices in adjacent cells to output a final quad mesh. We made this algorithmic choice in the interest of simplicity, and to draw a clear analogy between our method and the original work by \citet{Ju2002}. However, our work could likely benefit from orthogonal-yet-complementary improvements to the original Dual Contouring, e.g., by enforcing manifold outputs \cite{schaefer2007manifold}, which would not require major fundamental changes to our method. More interestingly, our algorithm's outer loop relies on triangulating the prior iteration's quads to build the next optimization problem. While we followed an arbitrary strategy (taking the first diagonal), future work may consider using the SDF data to optimize the choice of triangulation itself as suggested recently by \citet{Shen2023}.
Similarly, while our implementation assumed signed distance samples to lie on a regular grid, future work may consider extensions of our method to adaptive data structures like octrees.

\begin{figure}
\centering
\includegraphics{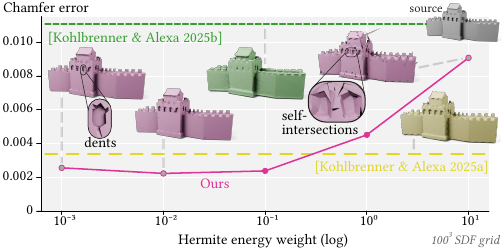}
\caption{Effect of the Hermite energy weight $\dcweight$ on the fidelity of our reconstruction. Large values make the method unstable and cause spikes to appear. We set a default value of $0.02$.}
\label{fig:ablation-dc-weight}
\end{figure}

Finally, this paper sits on the shoulders of recent work that has reawakened our community's interest in surface reconstruction from discrete SDF samples.
By focusing our study on the recovery of geometric sharp features, we have contributed to this growing area of research that has revealed itself to possess both impressive mathematical richness and practical importance.
As the literature on this topic continues to expand, we also hope to see future work dedicated to building datasets, baselines, and open challenges that can help guide SDF reconstruction towards a healthy maturity.

\begin{acks}
The Geometry and the City lab at Columbia University is supported by generous gifts from nTop, Adobe, Dandy, and Braid Technologies, as well as by a sponsored research project from Dreamsports and the Columbia Engineering Interdisciplinary Research Fund.
The third author acknowledges the generous support from the Natural Sciences and Engineering Research Council of Canada (Grant RGPIN-2021-02524).
The fourth author acknowledges the generous support from the National Science Foundation (award \#2335493) and a gift from Adobe.

We thank the authors of the 3D models used throughout this paper for making them available for academic use. Figures in this work contain the playing card box \cite{playing_card_box}, mechanical toy piece \cite{tardis_transformer}, metatron \cite{metratron-mesh}, Empire State \cite{empire_state}, Great Wall of China \cite{great_wall}, dragon head \cite{rathalos_head}, bunny \cite{bunny-mesh}, sword \cite{buster_sword}, dualstrusion ball in cube \cite{dualstrusion}, Chichen Itza \cite{pyramid-mesh}, and car piece \cite{openrc_car}.
\end{acks}

\begin{figure*}[]
\centering
\includegraphics[width=0.93\textwidth]{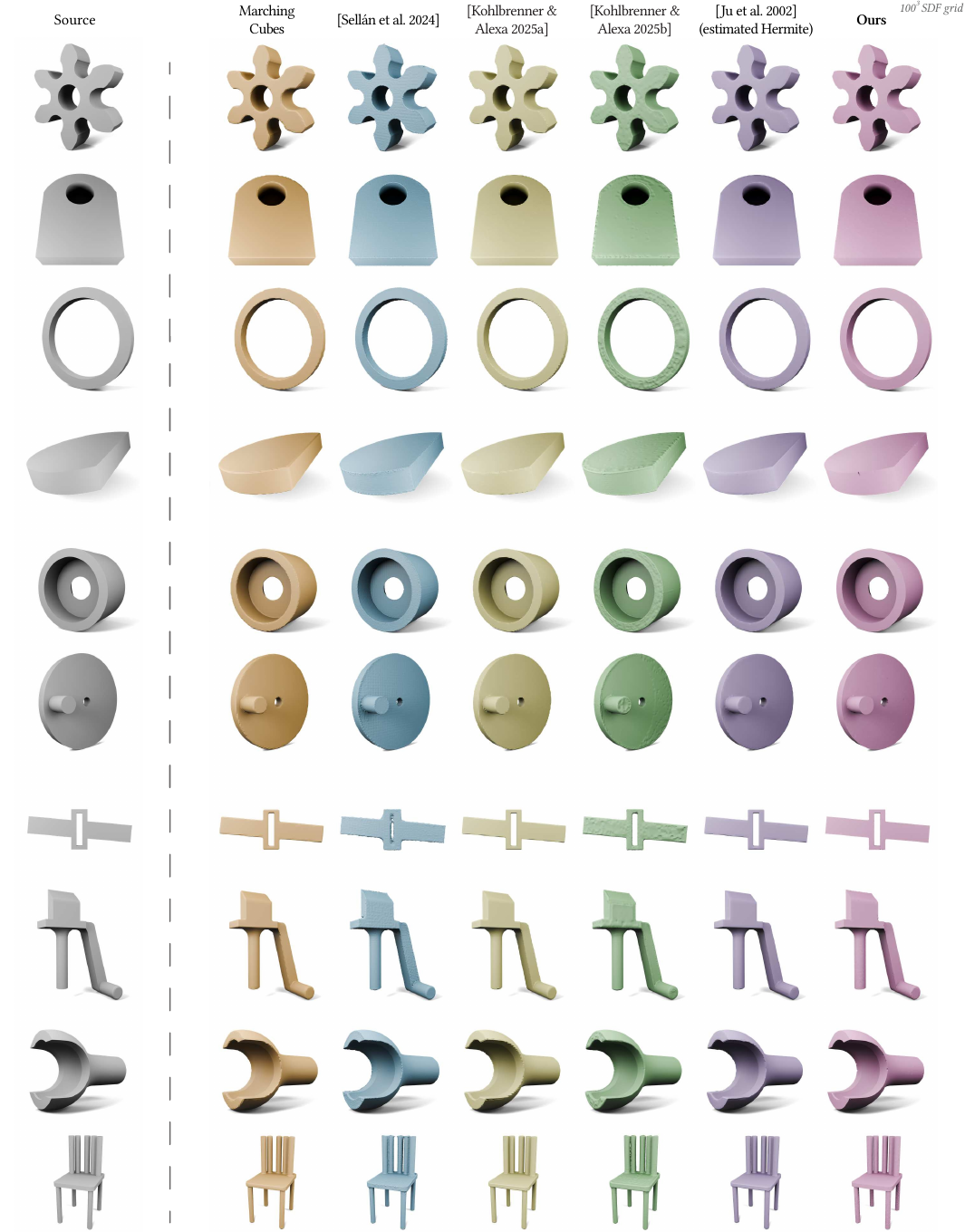}
\caption{A comparison against other methods on a subset of the ABC dataset, at a resolution of $100^3$.}
\label{fig:large-scale-test}
\end{figure*}

\bibliographystyle{ACM-Reference-Format}
\bibliography{bibliography}

\clearpage

\appendix
\setcounter{page}{1}

\section{Pseudocode}

{\SetAlgoNoLine%
{\small
\setlength{\textfloatsep}{2pt}
\setlength{\intextsep}{2pt}
\begin{algorithm}
\caption{Dual Contouring of signed distance data}
\label{alg:pseudocode}

\Fn{\SDFDualContouring{$\cell_1, \dots, \cell_m,
\cellcorner_1, \dots, \cellcorner_n,
\sdf_1, \dots, \sdf_n$}}{

    Identify interesting edges and cells using signs of $\sdf_i$

    \ForEach{interesting edge $\edge$}{
        Estimate $\Hermitepoint^0,\Hermitenormal^0$ using
        Equations~\ref{equ:Hermitepointest}, \ref{equ:Hermitenormalest}
    }

    \ForEach{interesting cell $\cell_i$}{
        $\centroid_i \gets$ average of Hermite data contained in $\cell_i$

        $\cellvert_i^0 \gets \centroid_i$
    }

    \For{$k = 0,\dots,\text{max\_outer\_iter}$}{

        $\mesh^k \gets$ connect vertices to form global mesh

        $\fip_1,\dots \gets$ intersect edges of $\mesh^k$ with cell faces to find
        face intersection points

        Triangulate global mesh $\mesh^k$

        \ForEach{SDF data point $(\cellcorner_i,\sdf_i)$}{
            Find closest point to $\cellcorner_i$ on $\mesh^k$

            Find cell $c_j$ closest point belongs to

            Assign $(\cellcorner_i,\sdf_i)$ to $c_j$
        }

        \ForEach(\tcp*[f]{parallel}){cell $\cell_i$}{
            $\cellvert^{k,0}\gets\cellvert^{k}$

            \For{$r=1,\dots,\text{max\_inner\_iters}$}{
                $\mesh_i^{k,r}\gets$ build local mesh using $\cellvert^{k,r}$

                $\t^r_1,\q^r_1,\dots\gets$
                compute closest points on local mesh to assigned $\cellcorner$

                $\cellvert^{k,r+1} \gets$ solve quadratic minimization problem

                \lIf{$\|\cellvert^{k,r+1} - \cellvert^{k,r}\|\leq \text{tol}$}{
                    \KwBreak
                }
            }

            $\cellvert^{k+1} \gets \cellvert^{k,r_{\mathrm{last}}}$
        }

        \ForEach{interesting edge}{
            $\Hermitepoint^{k+1},\Hermitenormal^{k+1}\gets$
            update Hermite data using \autoref{equ:Hermiteupdate}
        }
    }

    $\mesh\gets$ connect vertices to form final global mesh

    \Return{$\mesh$}
}

\end{algorithm}}%

\section{Tables}

Tables \ref{tab:quant_results_noise}, \ref{tab:quant_results} and \ref{tab:quant_results_10shapes} show numerical results for the large-scale quantitative results explained in \autoref{sec:results}.

\newcolumntype{C}[1]{>{\centering\arraybackslash\hspace{0pt}}p{#1}}

\begin{table}
\caption{Quantitative comparison under increasing noise (averaged over 10 shapes, at resolution $100^3$). Best scores are bold; second best are underlined.}
\begin{center}
\rowcolors{2}{gray!10}{white}
\resizebox{\columnwidth}{!}{%
\begin{tabular}{
  C{0.22\textwidth}||C{0.13\textwidth}|C{0.13\textwidth}|C{0.14\textwidth}
}
\toprule
\rowcolor{headercolor}
Method & Chamfer error ($\times 10^{-3}$) & Hausdorff error ($\times 10^{-2}$) & Edge chamfer error \\
\midrule
\rowcolor{header2color}
\multicolumn{4}{c}{Noise = 0.001} \\
\midrule
\cite{Lorensen1987} & 4.552 & 2.448 & 0.2272 \\
\cite{Ju2002} (est. Hermite)& 3.698 & \underline{2.129} & \underline{0.1253} \\
\cite{kohlbrenner2025isosurface} & \underline{3.611} & \textbf{2.107} & 0.1468 \\
\cite{kohlbrenner2025polyhedral} & 4.808 & 2.448 & 0.3137 \\
Ours & \textbf{2.683} & 2.651 & \textbf{0.04688} \\
\midrule
\rowcolor{header2color}
\multicolumn{4}{c}{Noise = 0.005} \\
\midrule
\cite{Lorensen1987} & 6.338 & \underline{2.117} & \textbf{0.08841} \\
\cite{Ju2002} (est. Hermite)& 6.543 & 5.037 & \underline{0.1022} \\
\cite{kohlbrenner2025isosurface} & \textbf{3.612} & \textbf{2.109} & 0.1401 \\
\cite{kohlbrenner2025polyhedral} & 9.734 & 3.3 & 0.2394 \\
Ours & \underline{5.104} & 4.816 & 0.1027 \\
\midrule
\rowcolor{header2color}
\multicolumn{4}{c}{Noise = 0.01} \\
\midrule
\cite{Lorensen1987} & 11.8 & \underline{3.713} & 0.1147 \\
\cite{Ju2002} (est. Hermite)& 12.8 & 6.927 & \textbf{0.1124} \\
\cite{kohlbrenner2025isosurface} & \textbf{3.614} & \textbf{2.11} & 0.1423 \\
\cite{kohlbrenner2025polyhedral} & 24.32 & 4.093 & 0.2183 \\
Ours & \underline{10.79} & 8.729 & \underline{0.1138} \\
\bottomrule
\end{tabular}
}
\end{center}
\label{tab:quant_results_noise}
\end{table}

\begin{table*}
\caption{A quantitative comparison of error metrics (averaged over 55 random shapes from the ABC dataset), across three different resolutions. Best scores are bold; second best are underlined. Our method yields the best result in all but two instances.}
\begin{center}
\renewcommand{\arraystretch}{0.9}
\rowcolors{2}{gray!10}{white}
\begin{tabular}{
  C{0.22\textwidth}||C{0.13\textwidth}|C{0.13\textwidth}|C{0.14\textwidth}|C{0.13\textwidth}|C{0.13\textwidth}
}
\rowcolor{headercolor}
Method & Chamfer error ($\times 10^{-3}$) & Hausdorff error ($\times 10^{-2}$) & Edge chamfer error & SDF energy ($\times 10^{-5}$) & \#Vertices \\

\midrule
\rowcolor{header2color}
\multicolumn{6}{c}{Resolution 50$^3$} \\
\midrule
\cite{Lorensen1987} & 7.586 & 3.692 & 0.3636 & 13.34 & 3.4k \\
\cite{Ju2002} (est. Hermite)& \underline{4.71} & \underline{2.937} & 0.356 & 8.611 & \textbf{3.4k} \\
\cite{rfta} & 4.976 & \textbf{2.88} & \underline{0.1069} & \textbf{0.4781} & 151k \\
\cite{kohlbrenner2025isosurface} & 5.063 & 3.033 & 0.2996 & 8.528 & 27k \\
\cite{kohlbrenner2025polyhedral} & 13.04 & 3.517 & 0.4066 & 3.826 & 73k \\
Ours & \textbf{2.958} & 2.97 & \textbf{0.03787} & \underline{1.074} & \textbf{3.4k} \\
\midrule
\rowcolor{header2color}
\multicolumn{6}{c}{Resolution 100$^3$} \\
\midrule
\cite{Lorensen1987} & 2.62 & 1.81 & 0.4171 & 3.061 & \textbf{14k} \\
\cite{Ju2002} (est. Hermite)& \underline{1.589} & \underline{1.372} & 0.3502 & 1.866 & \textbf{14k} \\
\cite{rfta} & 30.8 & 6.284 & \underline{0.1143} & 130.8 & 274k \\
\cite{kohlbrenner2025isosurface} & 1.843 & 1.492 & 0.274 & \underline{1.812} & 123k \\
\cite{kohlbrenner2025polyhedral} & 7.242 & 2.216 & 0.4466 & 2.321 & 163k \\
Ours & \textbf{0.7788} & \textbf{1.221} & \textbf{0.02622} & \textbf{0.05776} & \textbf{14k} \\
\midrule
\rowcolor{header2color}
\multicolumn{6}{c}{Resolution 150$^3$} \\
\midrule
\cite{Lorensen1987} & 1.329 & 1.107 & 0.3256 & 1.175 & 32k \\
\cite{Ju2002} (est. Hermite)& \underline{0.8344} & \underline{0.8513} & 0.3139 & 0.7912 & \textbf{32k} \\
\cite{kohlbrenner2025isosurface} & 1.118 & 0.9435 & 0.2608 & \underline{0.7778} & 281k \\
Ours & \textbf{0.4432} & \textbf{0.7037} & \textbf{0.01684} & \textbf{0.241} & \textbf{32k} \\
\midrule
\rowcolor{header2color}
\multicolumn{6}{c}{Resolution 200$^3$} \\
\midrule
\cite{Lorensen1987} & 1.059 & 0.9969 & 0.3003 & 1.115 & 57k \\
\cite{Ju2002} (est. Hermite)& \underline{0.7411} & \underline{0.8208} & 0.2929 & 0.9148 & \textbf{57k} \\
\cite{kohlbrenner2025isosurface} & 1.037 & 0.8769 & \underline{0.1999} & \underline{0.9096} & 496k \\
Ours & \textbf{0.4576} & \textbf{0.6741} & \textbf{0.01599} & \textbf{0.4359} & \textbf{57k} \\
\bottomrule
\end{tabular}
\end{center}
\label{tab:quant_results}
\end{table*}

\begin{table*}
\caption{Results from our method and related work on 10 smooth shapes used for evaluation by \citet{rfta}.}
\begin{center}
\renewcommand{\arraystretch}{0.9}
\rowcolors{2}{gray!10}{white}
\begin{tabular}{
  C{0.22\textwidth}||C{0.13\textwidth}|C{0.13\textwidth}|C{0.14\textwidth}|C{0.13\textwidth}|C{0.13\textwidth}
}
\toprule
\rowcolor{headercolor}
Method & Chamfer error ($\times 10^{-3}$) & Hausdorff error ($\times 10^{-2}$) & Edge chamfer error & SDF energy ($\times 10^{-5}$) & \#Vertices \\
\midrule
\rowcolor{header2color}
\multicolumn{6}{c}{Resolution 50$^3$} \\
\midrule
\cite{Lorensen1987} & 11.47 & 8.033 & 0.1897 & 29.06 & 2.5k \\
\cite{Ju2002} (est. Hermite)& 10.23 & 8.803 & 0.1842 & 28.75 & \textbf{2.5k} \\
\cite{rfta} & 11.35 & \textbf{5.433} & 0.2002 & \textbf{0.1906} & 109k \\
\cite{kohlbrenner2025isosurface} & \underline{9.813} & 7.648 & \textbf{0.1403} & 25.79 & 23k \\
\cite{kohlbrenner2025polyhedral} & 13.27 & \underline{5.6} & 0.2919 & \underline{0.3155} & 97k \\
Ours & \textbf{9.479} & 6.58 & \underline{0.1601} & 9.801 & \textbf{2.5k} \\
\midrule
\rowcolor{header2color}
\multicolumn{6}{c}{Resolution 100$^3$} \\
\midrule
\cite{Lorensen1987} & 4.579 & 5.233 & 0.1582 & 9.785 & 11k \\
\cite{Ju2002} (est. Hermite)& 4.206 & 6.402 & \underline{0.1402} & 9.55 & \textbf{11k} \\
\cite{rfta} & 6.039 & \underline{4.698} & 0.1696 & \textbf{0.1275} & 201k \\
\cite{kohlbrenner2025isosurface} & \underline{3.852} & 5.224 & \textbf{0.1092} & 8.721 & 121k \\
\cite{kohlbrenner2025polyhedral} & 5.445 & \textbf{3.444} & 0.3845 & \underline{0.1496} & 189k \\
Ours & \textbf{3.804} & 4.745 & 0.1457 & 6.346 & \textbf{11k} \\
\midrule
\rowcolor{header2color}
\multicolumn{6}{c}{Resolution 150$^3$} \\
\midrule
\cite{Lorensen1987} & 2.99 & 4.375 & 0.1462 & 7.618 & 25k \\
\cite{Ju2002} (est. Hermite)& \underline{2.681} & 5.413 & 0.1245 & 7.51 & \textbf{25k} \\
\cite{kohlbrenner2025isosurface} & \textbf{2.477} & \underline{4.241} & \underline{0.1236} & \underline{7.324} & 306k \\
Ours & 2.811 & \textbf{3.95} & \textbf{0.1064} & \textbf{6.072} & \textbf{25k} \\
\bottomrule
\end{tabular}
\end{center}
\label{tab:quant_results_10shapes}
\end{table*}

\section{Quadratic minimization}

We now give additional details on how we solve the minimization problem in the inner loop. At a given inner iteration $r$ of a fixed outer iteration, we optimize the position of the cell vertex by solving
\begin{equation}
  \cellvert^{r + 1} = \argmin_{\cellvert} \dcweight^2 \sum_{e_j \in c_i} \left( (\cellvert - \Hermitepoint_j) \cdot \Hermitenormal_j \right)^2 + \| A_d^r \cellvert - b_d^r \|^2 + \mu\|\cellvert - \cellvert^{r}\|^2.
\end{equation}}
This is a quadratic minimization problem, which can be expressed in the form
\begin{equation}
    \cellvert^{r + 1} = \argmin_{\cellvert} \| Q \cellvert - c \|^2 = \argmin_{\cellvert} \cellvert^T Q^T Q \cellvert - 2 c^T Q \cellvert + c^T c,
\end{equation}
where $Q$ is a matrix of size $(N_H + N_d + 3) \times 3$ (with $N_H$ being the number of edges with Hermite data of cell $c_i$, and $N_d$, the number of SDF data points assigned to $c_i$), and $c$ is a vector of size $N_H + N_d + 3$.

In particular, $Q$ is built by vertically stacking the following three matrices:
\begin{equation}
    Q_{dc} = \dcweight \begin{bmatrix}
\Hermitenormal_1^T \\
\Hermitenormal_2^T \\
\vdots \\
\Hermitenormal_{N_H}^T
\end{bmatrix} \in \mathbb{R}^{N_H \times 3}, \quad Q_d = A_d^r \in \mathbb{R}^{N_d \times 3}, \quad Q_\mu = \sqrt{\mu} I_3 \in \mathbb{R}^{3 \times 3},
\end{equation}
where each row of $Q_d$ stores $\alpha \cdot \dvec^T$, denoting by $\alpha$ the barycentric coordinate of $\t^r_j$ with respect to $\cellvert^r$.
Similarly, $c$ is built by vertically stacking the following three vectors:
\begin{equation}
    c_{dc} = \dcweight \begin{bmatrix}
\Hermitenormal_1^T \Hermitepoint_1 \\
\Hermitenormal_2^T \Hermitepoint_2 \\
\vdots \\
\Hermitenormal_{N_H}^T \Hermitepoint_{N_H}
\end{bmatrix} \in \mathbb{R}^{N_H}, \quad c_d = b_d^r \in \mathbb{R}^{N_d}, \quad c_\mu = \sqrt{\mu} \cellvert^r \in \mathbb{R}^{3}.
\end{equation}
Each row of $c_d$ then stores $q_j \cdot \dvec_j - \beta \Hermitepoint \cdot \dvec_j - \gamma \fip \cdot \dvec_j$, where $\beta$ and $\gamma$ are the barycentric coordinates of $\t^r_j$ corresponding to $\h_j$ and $\fip_j$, respectively. When $\alpha = 0$, we set $\alpha = 1$, $\beta = 0$ and $\gamma = 0$, avoiding numerical locking by encouraging the vertex to move towards or away from the sphere.
Finally, we solve the system following the same approach as \citet{Ju2002} (see their Sec. 2.3).

\section{Additional results}
\begin{figure}[h]
\centering
\includegraphics{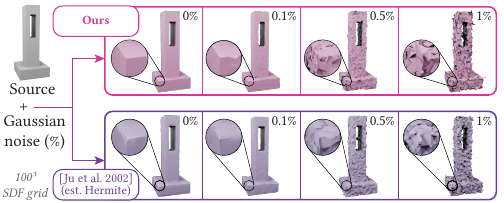}
\caption{Comparison of the results of our method and \cite{Ju2002} under different noise levels.}
\label{fig:noise}
\end{figure}

\end{document}